\documentclass[prd,onecolumn,showpacs,nofootinbib,preprintnumbers,superscriptaddress]{revtex4}
\usepackage{amsmath}
\usepackage{amsfonts}
\usepackage{graphicx}
\usepackage{hyperref}
\usepackage{mathbbol}

\def\be{\begin{equation}}
\def\ee{\end{equation}}
\def\bea{\begin{eqnarray}}
\def\eea{\end{eqnarray}}
\def\f{\frac}
\def\s{\sqrt}
\def\l{\left}
\def\r{\right}
\def\e{\epsilon}

\def\k{\kappa}
\def\no{\nonumber}

\begin{document}
\title{Scalar field cosmology: its non-linear Schr\"{o}dinger-type formulation}
\date{\today}

\vspace{2cm}

\author{Burin Gumjudpai}
  \affiliation{Centre for
Theoretical Cosmology,
  DAMTP, University~of~Cambridge\\ CMS, Wilberforce Road, Cambridge CB3
0WA, United Kingdom}\email{B.Gumjudpai@damtp.cam.ac.uk}
 \affiliation{Fundamental Physics \&
Cosmology Research Unit, The Tah Poe Academia Institute~(TPTP)\\
Department of Physics,~Naresuan University,~Phitsanulok 65000,~Siam}\email{buring@nu.ac.th}

\date{\today}

\begin{abstract}
Scalar field cosmology is a model for dark energy and inflation. It
has been recently found that the standard Friedmann formulation of
the scalar field cosmology can be expressed in a non-linear
Schr\"{o}dinger-type equation. The new mathematical formulation is
hence called non-linear Schr\"{o}dinger (NLS) formulation which is
suitable for a FRLW cosmological system with non-negligible
barotropic fluid density. Its major features are reviewed here.
\end{abstract}


\date{\today}

\vskip 1pc

\maketitle \vskip 1pc
\section{Introduction}
\label{sec:introduction} The present universe is under accelerating
expansion. This is convinced by many present observational data from
cosmic microwave background \cite{Masi:2002hp}, large scale
structure surveys \cite{Scranton:2003in} and supernovae type Ia
\cite{Riess:1998cb, Riess:2004nr,Astier:2005qq}. There are many
ideas to explain such an expanding state, mainly it can be
classified into three types: braneworlds and modification of
gravitational theory (e.g. \cite{Dvali}), backreaction effect from
inhomogeneity \cite{Kolb:2005me} and dark energy (for review, see
\cite{Padmanabhan:2004av}). Dark energy is a type of cosmological
fluid appearing in the matter term of the Einstein equation with
equation of state $w_{\rm D.E.} < -1/3$ so that it can generate
repulsive gravity and therefore accelerating the universe. The
simplest dark energy model is just a cosmological constant with
$w_{\Lambda} = -1$. However the cosmological constant suffers from
fine-tuning problem. Observational data suggests that the present
value of $w_{\rm D.E.}$ is very close to -1 and it also allows
possibility that dark energy could be dynamical. Therefore scalar
field model of dark energy became interesting topic in cosmology
since time-evolving behavior of the scalar field gives hope for
resolving the fine-tuning problem. Although the scalar field has not
yet been observed, it is motivated from many ideas in high energy
physics and quantum gravities. Theoretical predictions of its
existence at TeV scale could be tested at LHC and Tevatron in very
near future. Phenomenologically the scalar field is also motivated
in model building of inflation where super-fast acceleration happens
in the early universe \cite{inflation}. Cosmic microwave background
data combined with other results allows possibility that the scalar
field could be phantom, i.e. having equation of state coefficient
$w_{\phi} < -1$ \cite{Hinshaw:2008kr}. The phantom equation of state
is attained from negative kinetic energy term in its Lagrangian
density \cite{Caldwell:1999ew, Melchiorri:2002ux}. The most recent
five-year WMAP result \cite{Dunkley:2008ie} combined with Baryon
Acoustic Oscillation of large scale structure survey from SDSS and
2dFGRS \cite{Percival:2007yw} and type Ia supernovae data from HST
\cite{Riess:2004nr}, SNLS \cite{Astier:2005qq} and ESSENCE
\cite{WoodVasey:2007jb} assuming dynamical $w$ with flat universe
yields $-1.38 < w_{\phi,0} < -0.86$ at 95\% CL and $w_{\phi,0} =
-1.12 \pm 0.13$ at 68\% CL. With additional BBN constraint of limit
of expansion rate \cite{Steigman:2007xt,Wright:2007vr}, $w_{\phi,0}
= -1.09 \pm 0.12$ at 68\% CL. The phantom field will finally
dominate the universe in future, leading to Big Rip singularity
\cite{Caldwell:2003vq}. There have been many attempts to resolve the
singularity from both phenomenological and fundamental inspirations
\cite{Sami:2005zc}. However fundamental physics of the phantom field
is still incomplete due to severe UV instability of the field's
quantum vacuum state \cite{carroll}.

 This review interests in non-linear
Schr\"{o}dinger-type formulation of scalar field cosmology. We shall
call the formulation, NLS formulation. In our NLS system,
cosmological ingredients are scalar field and a barotropic fluid
with constant equation of state, $p_{\gamma}
=w_{\gamma}\rho_{\gamma}$. We also have non-zero spatial curvature.
This is a system resembling of our present universe filled with
scalar field dark energy and barotropic cold dark matter or of the
early inflationary universe in presence of inflaton and other fields
behaving barotropic-like considered in e.g.
\cite{Chaicherdsakul:2006ui}. In such a model, the scale invariant
spectrum in the cosmic microwave background was claimed to be
generated not only from fluctuation of scalar field alone but rather
from both scalar field and interaction between gravity to other
gauge fields such as Dirac and gauge vector fields.

Not long ago, mathematical alternatives to the standard Friedmann canonical scalar field cosmology with barotropic perfect fluid, was proposed
e.g. non-linear Ermakov-Pinney equation \cite{Hawkins:2001zx, Williams:2005bp}. Expressing standard cosmology with $k>0$ in Ermakov equation
system yields a system similar to Bose-Einstein condensates \cite{Lidsey:2003ze}. Another example is a connection from a generalized
Ermakov-Pinney equation with perturbative scheme to a generalized WKB method of comparison equation \cite{Kamenshchik2006}. It was then realized
that solutions of the generalized Ermakov-Pinney equation are correspondent to solutions of a non-linear Schr\"{o}dinger-type equation, and then
the NLS version of the Friedmann-Lema\^{i}tre-Robertson-Walker (FLRW) cosmology was formulated \cite{Williams:2005bp}. In the NLS framework, the
system of FLRW cosmological equations: Friedmann equation, acceleration equation and fluid equation are written in a single non-linear
Schr\"{o}dinger-type equation. We will not prove it here but instead, referring to Ref. \cite{D'Ambroise:2006kg}. Few recent applications
\cite{Gumjudpai:2007bx,Gumjudpai:2007qq,Phetnora:2008,Gumjudpai:2008} of the NLS formulation have been made and this review intends to conclude
its major aspects.  Note that application of Schr\"{o}dinger-type equation to scalar field cosmology was previously made but in different form
to tackle aspects of inflation and phantom field  \cite{Chervon:1999}.

\section{Scalar field cosmology} \label{sec:NS}
\subsection{Friedman formulation}
We set up major concepts in this section before considering its
application later. In the Friedmann system, barotropic fluid has
pressure $p_{\gamma}$ and density $\rho_{\gamma}$ with an equation
of state, $p_{\gamma} = w_{\gamma}\rho_{\gamma}= [(n-3)/3]
\rho_{\gamma}$ where $ n = 3(1+w_{\gamma})$. Scalar field pressure
obeys $p_{\phi} = w_{\phi}\rho_{\phi}$. To sum up, $\rho_{\rm tot} =
\rho_{\gamma} + \rho_{\phi}$ and $p_{\rm tot} = p_{\gamma} +
p_{\phi}$. Therefore $n=0$ means $w_{\gamma}=-1$. The others are:
$n=2$ for $w_{\gamma}=-1/3$; $n=3$ for $w_{\gamma}=0$; $n=4$ for
$w_{\gamma}=1/3$; $n=6$ for $w_{\gamma}=1$. Barotropic fluid and
scalar fluid are conserved separately. Dynamics of the barotropic is
governed by fluid equation, $ \dot{\rho}_{\gamma} = - n H
\rho_{\gamma} $ with solution,
$\rho_{\gamma} = {D}/{a^{n}}\,,$
 where $a$ is scale factor. The dot denotes time derivative. $H =
\dot{a}/a$ is Hubble parameter and $D\geq 0$ is a proportional
constant. Scalar field is minimally-coupled to gravity with
Lagrangian density, $ \mathcal{L} = (1/2)\epsilon \dot{\phi}^2 -
V(\phi)$ and is homogenously spread all over the universe. The
scalar field density and pressure are
\bea \rho_{\phi} = \frac{1}{2} \epsilon \dot{\phi}^2 + V(\phi)\,,
\;\;\;\;\;\;\;\;
 p_{\phi}= \frac{1}{2} \epsilon \dot{\phi}^2 - V(\phi)\,.
\label{phanp}\eea The branch $\epsilon=1$ is for non-phantom field
case and $\epsilon=-1$ is for phantom field case. Dynamics of the
scalar field is controlled by conservation equation,
$\,\epsilon(\ddot{\phi} + 3H\dot{\phi}) = -{{\rm d}V}/{{\rm
d}\phi}\,,$
in which the spatial expansion $H$ of the universe sources friction
to
dynamics of the field. The Hubble parameter is governed by Friedmann equation, %
$ H^2 = ({\kappa^2}/{3})\rho_{\rm tot} - {k}/{a^2}\,, $ and by
acceleration equation,
$ {\ddot{a}}/{a} =  -({\kappa^2}/{6}) (\rho_{\rm tot}
 + 3p_{\rm tot})\,, $ 
which gives acceleration condition
  $
  p_{\rm tot} < -{\rho_{\rm tot}}/{3} \,.\label{acptot}
  $
Here $p_{\rm tot}= w_{\rm eff}\rho_{\rm tot}, \kappa^2 \equiv 8\pi G
= 1/M_{\rm P}^2$. $G$ is Newton's gravitational constant. $M_{\rm
P}$ is reduced Planck mass. $k$ is
spatial curvature and %
\bea w_{\rm eff} = \frac{{\rho_{\phi}w_{\phi} + \rho_{\gamma}
w_{\gamma}}}{\rho_{\rm tot}}\,. \eea
If we express the field speed and the field potential in term of $a(t)$ and time derivative of $a(t)$, then%
\bea   \epsilon  \dot{\phi}(t)^2  =  -\frac{2}{\kappa^2} \left[
\dot{H} - \frac{k}{a^2}  \right] - \frac{n D}{3  a^n} \;\;\;\;\;{\rm
and}\;\;\;\;\; V(\phi) = \frac{3}{\kappa^2} \left[H^2 +
\frac{\dot{H}}{3} + \frac{2k}{3 a^2} \right] +
\left(\frac{n-6}{6}\right)
\frac{D}{a^n}\,. \label{Vgr} \eea %

\subsection{NLS formulation}
NLS formulation is a mathematical alternative to the standard
Friedmann formulation with hope that the new formulation might
suggest some new mathematical tackling to problems in scalar field
cosmology. In the NLS formulation, there is no such an analogous
equation to Friedmann equation or fluid equation. Instead both of
them combine in single non-linear Schr\"{o}dinger-type equation,
 \bea u''(x) + \left[E-P(x)\right]
u(x)
 = -\frac{nk}{2}u(x)^{(4-n)/n}\,.   \label{schroeq} \eea
The links to cosmology are valid as one defines NLS quantities
\cite{D'Ambroise:2006kg}, \bea
 u(x) \equiv a(t)^{-n/2},\;\;\;
 E \equiv  -\frac{\kappa^2 n^2}{12} D, \;\;\;
P(x) \equiv \frac{\kappa^2 n}{4}a(t)^{n} \epsilon \dot{\phi}(t)^2
\,. \label{E} \eea
where $'$ denotes ${\rm d}/{\rm d}x$. Independent variable $t$ is
scaled to NLS independent variable $x$ as $ x = \sigma(t), $ such
that
\bea \dot{x}(t)= u(x)\;\;\;\;{\rm and} \;\;\;\;  \phi(t) =
\psi(x)\,, \label{phitopsi}\eea
which gives $\epsilon \dot{\phi}(t)^2 = \epsilon \dot{x}^2\,
\psi'(x)^2$. Hence $ \epsilon\,\psi'(x)^2 = ({4}/{ \kappa^2
n})\,P(x)\,$, and
 \bea \psi(x) =
\pm\frac{2}{\kappa\sqrt{n}}\int{\sqrt{\frac{P(x)}{\epsilon}}}\,{\rm
d}x\,  \,.
\label{phitoPx} %
\eea
Inverse function $\psi^{-1}(x)$ exists for $P(x) \neq 0$ and $n
\neq 0$. In this circumstance, $ x(t) = \psi^{-1}\circ \phi(t)$ and
the scalar field potential, $V\circ \sigma^{-1}(x)$ and $\epsilon
\dot{\phi}(t)^2$ can be expressed in NLS formulation as
\bea \epsilon \dot{\phi}(x)^2  =  \frac{4}{\kappa^2 n}u u'' +
\frac{2 k}{\kappa^2} u^{4/n} + \frac{4E}{ \k^2 n}u^2\,, \eea and
\bea
    V(x) =
\frac{12}{\kappa^2 n^2}( u')^2 - \frac{2 P}{\kappa^2 n}u^2  +
\frac{12 E }{\kappa^2 n^2}u^2 + \frac{3 k }{\kappa^2}u^{4/n}\,.
\label{vt} \eea
The other quantities are \bea \rho_{\phi} &=&
\frac{12}{\kappa^2 n^2}( u')^2   + \frac{12 E }{\kappa^2 n^2}u^2 +
\frac{3 k }{\kappa^2}u^{4/n} \,,
\label{NLSrho_phi} \\
p_{\phi} &=& -\frac{12}{\kappa^2 n^2}( u')^2   +  \frac{4
P}{\kappa^2 n}u^2 -\frac{12 E }{\kappa^2 n^2}u^2 - \frac{3 k
}{\kappa^2}u^{4/n} \,, \label{NLSrho_p} \\
  \rho_{\rm tot} & = & \frac{12}{\kappa^2 n^2}( u')^2   +
 \frac{3 k }{\kappa^2}u^{4/n} \,, \label{rhoNLS} \\
     p_{\rm tot} & = & -\frac{12}{\kappa^2 n^2}( u')^2   +\f{4 }{\k^2 n}u u'' -
 \frac{ k }{\kappa^2}u^{4/n}. \label{pNLS}\\
 H &=& -\f{2}{n} u' \,,\;\;\; \;\;\;\;\;\;\;\;\;\;\;\,\;\;\;\;\;\; \dot{H} =
-\f{2}{n} u u''\,, \label{NLSH} \\
 \ddot{\phi} &=&  \frac{2 P u u' + P' u^2}{\k \s{P \e n}}
\,, \;\;\;3H\dot{\phi} = - \frac{12 u' u }{n \k} \s{\frac{P}{\e n
}}\,. \label{ddotphiNLS}
 \eea
We shall see later examples that the program of NLS formulation must
start from presuming the ``wave function", $u(x)\equiv a^{-n/2} =
\dot{x}(t)$, before proceeding to calculate the other quantities.
We know that normalization condition for a wave function is %
$ \int_{-\infty}^{\infty}|u(x)|^2 {\rm d}x = 1\,.$ If applying this
to our NLS wave function, then $ \int_{-\infty}^{\infty} \dot{x}^2
{\rm d}x = 1\,$. In order to satisfy the condition, $x$ must be
constant (hence so is $t$) with an integrating constant = 1.  In
connecting Friedmann formulation to NLS formulation, we are forced
to have $u(x) = \dot{x}(t)$. Therefore $u(x)$ is, in general,
non-normalizable.

\section{Slow-roll conditions} \label{sec:slowroll}
\subsection{Slow-roll conditions: flat geometry and scalar field domination}
In flat universe with scalar field domination,
$ \dot{H} = -({\kappa^2}/2)\dot{\phi}^2\epsilon\,. $
Hence for $\epsilon = -1$ (phantom field), %
\be%
 0 < aH^2 < \ddot{a}\,, %
\ee %
i.e. the acceleration is greater than speed of expansion per Hubble
radius, $\dot{a}/cH^{-1}$ and for $\epsilon =1$ (non-phantom field),
\be%
 0  < \ddot{a} < aH^2 \,. %
\ee %
Slow-roll condition \cite{Liddle:1992wi,liddlebook} assumes
negligible kinetic term, i.e. $|\epsilon\dot{\phi}^2/2| \ll V(\phi)$
which makes an approximation $H^2 \simeq \kappa^2V/3$.
 This results in a condition $ |\dot{H}| \ll H^2$. Slow-roll parameter, $\varepsilon \equiv
-\dot{H}/H^2 $ is hence defined from this relation. The condition
$|\epsilon\dot{\phi}^2/2| \ll V(\phi)$ is then equivalent to
$|\varepsilon| \ll 1$, i.e.
 $-1 \ll \varepsilon < 0 $ for phantom field case and $0 <
\varepsilon \ll 1 $ for non-phantom field case. Considering $
\dot{H} \simeq 0 $ implying approximative constancy in $H$ during
the slow-rolling regime. For non-phantom field, this condition is
necessary for inflation to happen (though not sufficient)
\cite{Liddle:1992wi} however, for phantom field case, the negative
kinetic term always results in acceleration with $w_{\phi} \leq -1 $
then it does not need the slow-roll approximation. Another slow-roll
parameter can be defined when the friction term dominates
$|\ddot{\phi}| \ll |3H\dot{\phi}|$. This gives the second parameter,
$\eta \equiv -\ddot{\phi}/H\dot{\phi}$ and the approximation is made
to $|\eta| \ll 1$ \cite{Liddle:1992wi}. The field fluid equation is
then %
$ \dot{\phi} \simeq - {V_{\phi}}/{3\epsilon H}\, %
$ %
which implies that if $\epsilon = -1$, the field can roll up the
hill. With all assumptions imposed here, i.e. $k=0$,
$\rho_{\gamma}=0$, $|\epsilon\dot{\phi}^2/2| \ll V$ and
$|\ddot{\phi}| \ll |3H\dot{\phi}|$, one can derive %
$ \varepsilon = ({1}/{2\kappa^2
\epsilon})\left({V_{\phi}}/{V}\right)^2$  and $ \eta =
({1}/{\kappa^2}) ({V_{\phi\phi}}/{V})$ as known where the subscript
$_{\phi}$ denotes ${\rm d}/{\rm d}\phi$.

\subsection{Slow-roll conditions: non-flat geometry and non-negligible barotropic density}
There are also inflationary models in presence of other field
behaving barotropic-like apart from having only single scalar fluid
\cite{Chaicherdsakul:2006ui}. The the scale invariant spectrum in
the cosmic microwave background was claimed to be generated not only
from fluctuation of scalar field alone but rather from both scalar
field and interaction between gravity to other gauge fields.
Assuming this scenario with
  $k \neq 0$ and $\rho_{\gamma} \neq 0$, then %
\be %
\dot{H} = - \frac{\k^2}{2}\dot{\phi}^2 \e + \f{k}{a^2} - \f{n \k^2
}{6} \f{D}{a^n} \,. \label{coffeeMixdot} \ee The slow-roll condition
becomes $|\k^2 \epsilon\dot{\phi}^2/6| \ll (\k^2
V/3) - (k/a^2) + (\k^2 D/3 a^n)$ hence  %
\be %
H^2 \simeq -\frac{\dot{H}}{3} + \f{k}{3a^2} - \f{n \k^2}{18}\f{D}{a^n} + H^2 \,, %
\ee implying $  |-({\dot{H}}/{3}) + ({k}/{3a^2}) - ({n \k^2
D}/{18 a^n}) | \ll H^2 $. We can reexpress this slow-roll condition as %
\bea %
|\varepsilon + \varepsilon_k + \varepsilon_D| \; \ll \; 1\,,
\label{SUMep} \eea where $\varepsilon_k \equiv k/a^2 H^2$ and
$\varepsilon_D \equiv -n \k^2 D/6 a^n H^2 $. Another slow-roll
parameter $\eta$ is defined as $\eta \equiv
-\ddot{\phi}/H\dot{\phi}$, i.e. the same as the flat scalar field
dominated case since the condition $|\ddot{\phi}| \ll
|3H\dot{\phi}|$ is independent of $k$ and $\rho_{\gamma}$.

Writing the condition $|\epsilon\dot{\phi}^2/2| \ll V$ in NLS form
using Eqs. (\ref{E}) and (\ref{vt}),  \bea |P(x)| \ll
\frac{3}{n}\left[ \l( \f{u'}{u}\r)^2 + E \right] +
\f{3}{4}k\, n\, u^{(4-2n)/n}\,. \label{NLSa} %
\eea If the absolute sign is not used, the condition is then
$\epsilon\dot{\phi}^2/2 \ll V $, allowing fast-roll negative kinetic
energy. Then Eq. (\ref{NLSa}), when combined with the
NLS equation (\ref{schroeq}), yields %
\bea u''\ll \f{3}{n}\f{u'\,^2}{u} + \l( \f{3}{n}-1 \r)E u +
\frac{kn}{4} u^{(4-n)/n}\,.\eea Friedmann analog of this condition
can be obtained simply by using Eq. (\ref{Vgr}) in the condition.
Using Eq. (\ref{ddotphiNLS}), the second slow-roll condition, $|\ddot{\phi}| \ll |3H\dot{\phi}| $ in the NLS form is written as,  %
\be%
 \l| \f{P'}{P} \r|\: \ll \: \l|-2 \l(\f{6+n}{n}\r)  \frac{u'}{u} \r|\,. %
\ee This condition yields the approximation $3H \e \dot{\phi}^2
\simeq -{\rm d }V/{\rm d} \phi$ which, in NLS form, is %
\bea %
\frac{P'}{P} \: \simeq \:  - \frac{2u'}{u} \:= \: n H a^{n/2}\,.
\eea The slow-roll parameters $\varepsilon$, $\varepsilon_k$ and
$\varepsilon_D$, in NLS form, are
\bea \varepsilon = \f{n u u''}{2
{u'}^2} \,, \;\;\;\; \varepsilon_k  = \f{n^2 k u^{4/n}}{{4 u'}^2}
\,, \;\;\;\; \varepsilon_D   =  \f{n E}{2}\l(\f{u}{u'}\r)^2 \,,
\label{NLSslowroll}
 \eea
 therefore \bea \varepsilon_{\rm tot} = \varepsilon + \varepsilon_k
+ \varepsilon_D = \f{n}{2}\l(\f{u}{u'} \r)^2 P(x)\,. \eea Hence the
slow-roll condition, $|\varepsilon_{\rm tot}| \ll 1$, is just \bea
\l| \l(\f{u}{u'} \r)^2 P(x) \r| \ll 1\,. \label{NLSepcon} \eea
Another slow-roll parameter $\eta = -\ddot{\phi}/H\dot{\phi}$ can be
found as follow. First considering $\psi(x) = \phi(t)$ and Eq.
(\ref{phitoPx}), using relation ${\rm d}/{\rm d}t = \dot{x}\,{\rm
d}/{\rm d} x$, one can obtain \be \eta = \f{n}{2} \l(\f{u}{u'}
\f{\psi''}{\psi'} +1  \r) = \f{n}{2} \l(\f{u}{u'} \f{P'}{2P} +1
\r)\,. \label{NLSeta} \ee The slow-roll condition $|\eta| \ll 1$ in
NLS form is just \be \l|  \f{u}{u'} \f{P'}{2P} +1 \r| \, \ll \, 1\,.
\label{NLSetacon} \ee

\section{Acceleration condition} \label{sec:acc}
For the phantom field, since its kinetic term is always negative and
could take any large negative values, the slow-roll condition is not
needed. The acceleration equation is taken as acceleration condition
straightforwardly, i.e. $\ddot{a}
> 0 $  hence
\bea %
\epsilon \dot{\phi}(x)^2 &<& - \left(\frac{n-2}{2}\right) \frac{D}{a^n} + V\,. \label{accongr} %
\eea This, in NLS-type form, is equivalent to
\bea %
E-P &>& - \frac{2}{n} \left(\frac{u'}{u}\right)^2    - \frac{nk}{2}\left(\frac{u^{2/n}}{u}\right)^2 \,, \label{acconNLS} %
\eea which is reduced to \bea u'' < \frac{2}{n} \frac{{u'}^2}{u}\,.
\label{acconNLS2} \eea
with help of the Eq. (\ref{schroeq}). %
Using Eqs. (\ref{NLSH}), the acceleration condition is just
 $\varepsilon < 1$.

\section{Power-law cosmology} \label{sec:WKB}
\begin{figure}[t]
\begin{center}
\includegraphics[width=5.4cm,height=5.4cm,angle=0]{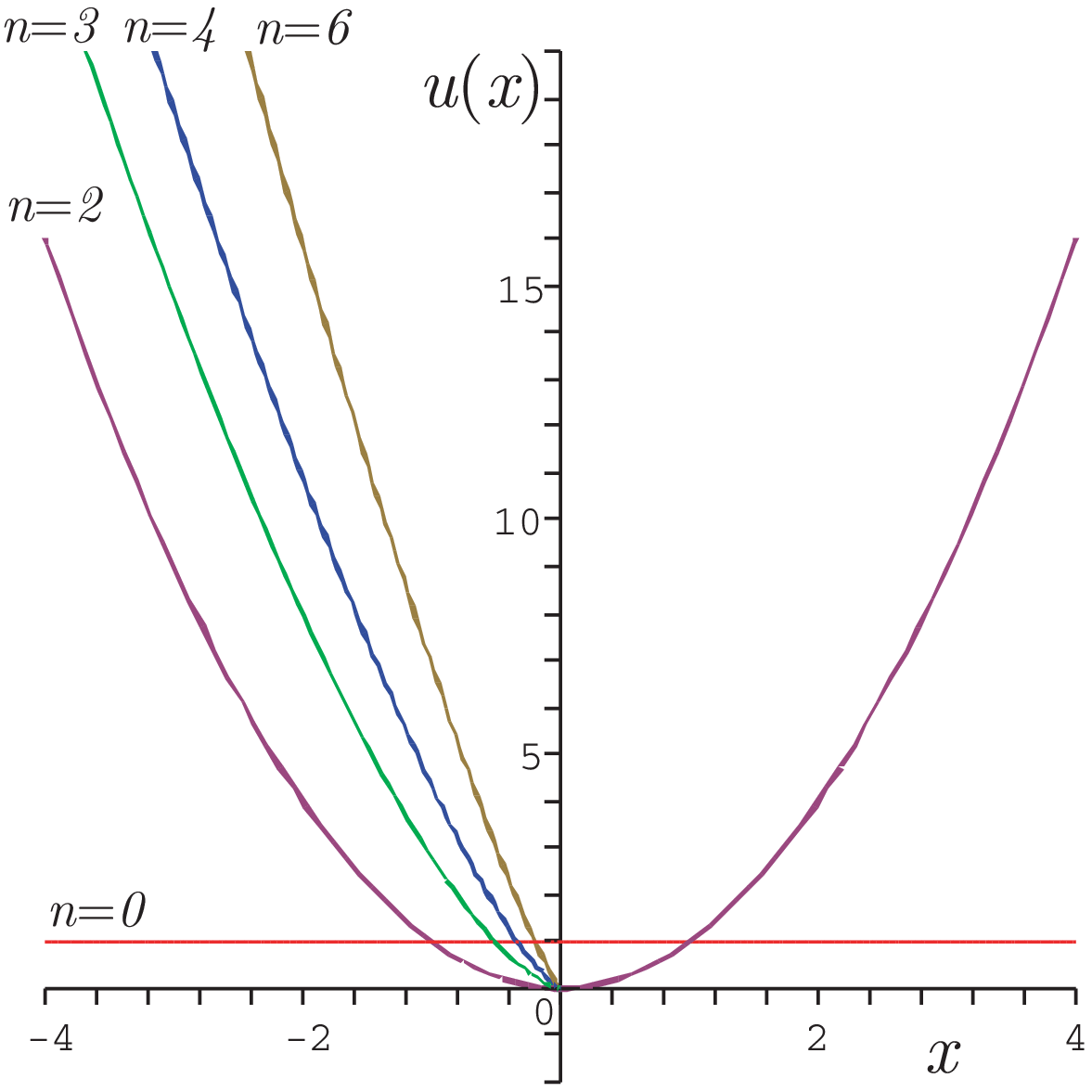}
\end{center}
\caption{$u(x)$ versus $x$ for power-law cosmology with $q=2$. We
set $x_0=0$. There is no real-value wave function for $n=3$, $n=4$
and $n=6$ unless $x<0$.} \label{pic5}
\begin{center}
\includegraphics[width=7.5cm,height=9.6cm,angle=0]{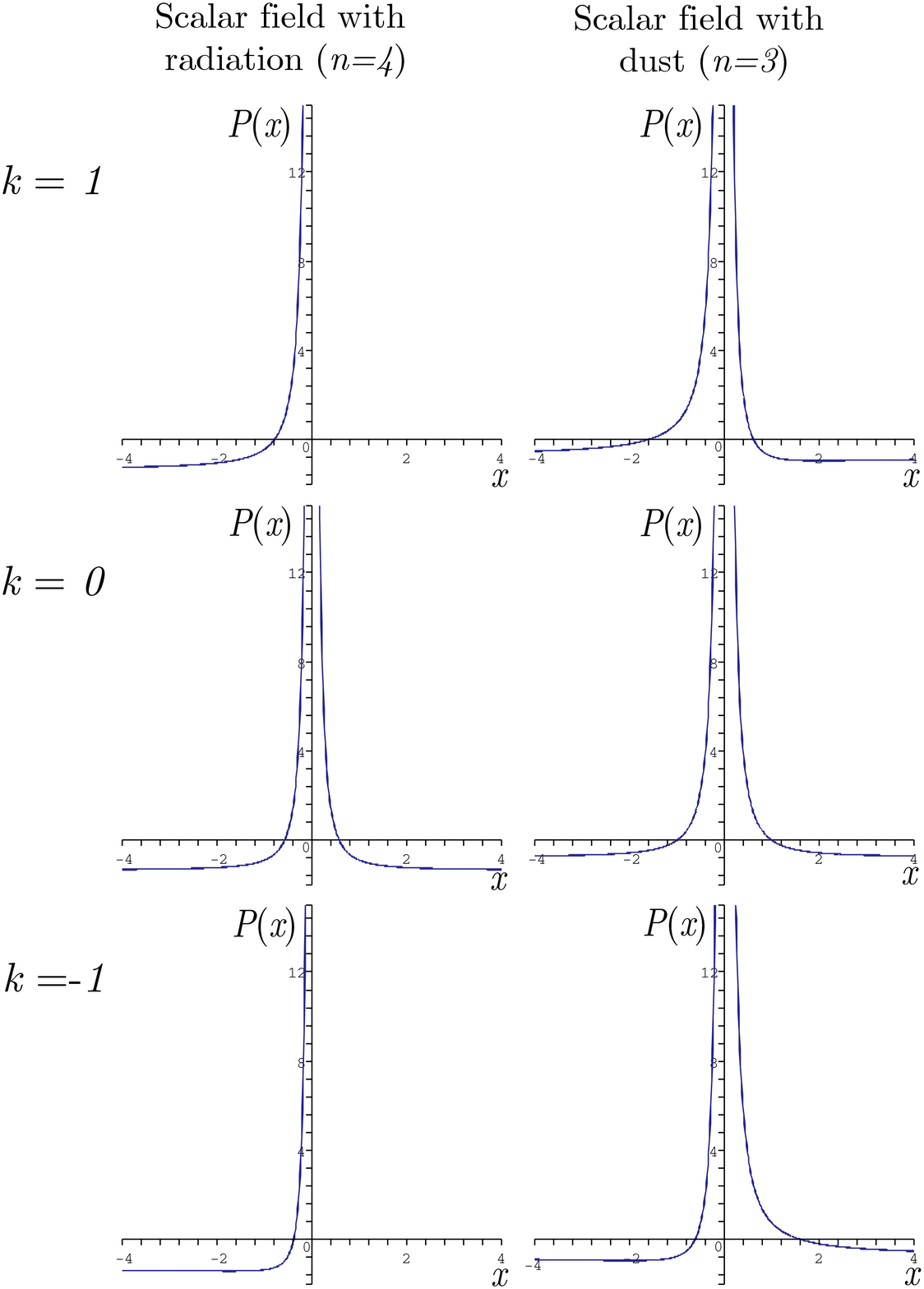}
\end{center}
\caption{$P(x)$ plotted versus $x$ for power-law expansion. We set
$q=2, \kappa=1, D=1 $ and $x_0=0$. There is only a real-value $P(x)$
for the cases $k=\pm 1$ with $n=4$ because, when $x>0$, $P(x)$
becomes imaginary in these cases. The physical value is when $x<0$
since $t$ has a reverse sign of $x$.} \label{pic4}
\end{figure}
The power-law expansion $a(t) = t^q\,,$  with $q > 1 $ is assumed
here as the first step of calculation. In some high-energy physics
models, during inflation, flat geometry and scalar field domination
are assumed. The universe was driven by an exponential potential
 $
V(\phi) = \left[{q(3q-1)}/({\kappa^2 t_0^2 })\right]
 \exp\left\{ -\kappa\sqrt{{2}/{q}}
 \left[\phi(t)-\phi(t_0)\right]\right\}\,\label{Vexp}
 $
\cite{Lucchin}. Also, at late time with dark matter component, the
expansion could be power-law. Recent results from X-Ray gas of
galaxy clusters put a constraint of $q \sim 2.3$ for $k = 0$, $q
\sim 1.14$ for $k = -1$ and $q \sim 0.95$ for $k = 1$
\cite{Zhu:2007tm}. For a flat universe, the power law expansion, $a=
t^q$, is attained when $-1< w_{\rm eff} < -1/3$ where $q =
2/[3(1+w_{\rm eff})]$. If using $q=2.3$ as above, it gives $w_{\rm
eff} = -0.71$ (only flat case). Latest combined WMAP5 results with
SNI and BAO yield $-0.0175 < \Omega_k < 0.0085$ at 95\% maximum
likelihood \cite{Dunkley:2008ie}. The mean is $\bar{\Omega}_k =
-0.0045$ corresponding to closed universe with $q= 0.986$
\cite{tepsuriya:2008}. Assuming power-law expansion,
the Schr\"{o}dinger wave function is \cite{Gumjudpai:2007qq} %
\be
u(x) = \dot{x}(t) = t^{-qn/2}   \,. \label{u(x)} \ee Integrating the
equation above so that the Schr\"{o}dinger scale, $x$
is related to cosmic time scale, $t$ as %
\be x = x(t) = -\frac{t^{-\beta }}{\beta} + x_0, \;\;\;\;{\rm and}
\;\;\;\; t(x) = \frac{1}{\left[-\beta (x-x_0) \right]^{1/\beta}}
\label{ttox}\,, \ee where $\beta \equiv (qn-2)/2 $ and $x_0$ is an
integrating constant. The parameters $x$ and $t$ have the same
dimension since $\beta$ is only a number. Then the wave function is
\be u(x) = \left[ \left( -\frac{1}{2}qn+1 \right)(x-x_0)
\right]^{qn/(qn-2)} \,, \label{uxt} \ee which depends on only $q$
and $n$. Wave functions for a range of barotropic fluids are
presented in Fig. \ref{pic5}. The result is confirmed by
substituting Eq. (\ref{uxt}) into Eq. (\ref{schroeq}). The field
speed and scalar potential are:
\be \epsilon\dot{\phi}(t)^2 = \frac{2q}{\kappa^2 t^2}  +
\frac{2k}{\kappa^2 t^{2q}} - \frac{n D}{3 t^{qn}} \;\;\;\; {\rm
and}\;\;\;\; V(t) = \frac{q(3q-1)}{\kappa^2 t^2} +
\frac{2k}{\kappa^2 t^{2q}} +
\left(\frac{n-6}{6}\right)\frac{D}{t^{qn}}\,. \label{vtpowerlaw} \ee
From Eq. (\ref{E}), therefore
the Schr\"{o}dinger potential is found to be %
\bea%
P(x) &=& \frac{2qn}{(qn-2)^2}\frac{1}{(x-x_0)^2}  +\:
\frac{kn}{2}\left[ \frac{-2}{(qn-2)(x-x_0)} \right]^{2q(n-2)/(qn-2)}
- \: \frac{\kappa^2 n^2 D }{12}\,. \label{Pxpower} \eea With $E =
-\kappa^2 n^2 D/12$, the Schr\"{o}dinger kinetic energy is
\bea T(x) &=&  - \: \frac{2qn}{(qn-2)^2}\frac{1}{(x-x_0)^2} -\:
\frac{kn}{2}\left[ \frac{-2}{(qn-2)(x-x_0)}
\right]^{2q(n-2)/(qn-2)}. \eea %
A disadvantage of Eq. (\ref{Pxpower}) is that we can not use it in
the case of scalar field domination. Dropping $D$ term in Eq.
(\ref{Pxpower}) can not be considered as scalar field domination
case since the barotropic fluid coefficient $n$ still appears in the
other terms. The non-linear Schr\"{o}dinger-type formulation is
therefore suitable when there are both scalar field and a barotropic
fluid together such as the situation when dark matter and scalar
field dark energy live together in the late universe or in the
inflationary models in presence of other fields behaving
barotropic-like and single scalar fluid
\cite{Chaicherdsakul:2006ui}.  $P(x)$ is plotted versus $x$ for
power-law expansion with $q=2$ in closed, flat and open universe in
Fig. \ref{pic4}. One can check that the acceleration condition
(\ref{acconNLS2}) for the power-law case is just $q >1$.

There is application of the NLS scalar field function $\psi$ in Eq.
(\ref{phitoPx}) to solve for scalar field exact solutions in
power-law, phantom expansion ($a\sim(t_{\rm a}-t)^q, q<0$) and
exponential (de Sitter) expansion $a \sim \exp(t/\tau)$
\cite{Gumjudpai:2007bx,Phetnora:2008}. For example in power-law
case: \bea \psi(x)  =  \frac{\pm 2}{\kappa \sqrt{n}} \times  \int
{\sqrt{ \f{2qn}{\e(qn-2)^2}\f{1}{(x-x_0)^2} + \f{kn}{2\e} \left[
\f{-2}{(qn-2)}\f{1}{(x-x_0)} \right]^{2q(n-2)/(qn-2)} -
\frac{\kappa^2n^2 D}{12 \epsilon}}}\: {\rm d}x  \,. \label{PowerPsi}
\eea %
The solution can be found only when assuming $k=0$,
\bea \phi(t) & =
& \pm \f{1}{qn-2} \sqrt{\f{2q}{\epsilon \kappa^2}}\left\{ \ln \left[
 \f{\,t^{-qn+2}}{\left(1 + \sqrt{1-(nD\kappa^2/6q)\,t^{-qn+2}}\right)^2}  \right]
+ 2\sqrt{1-\left(\f{nD\kappa^2}{6q}\right)\,t^{-qn+2}}  +
\ln\left(\f{qn-2}{2qn}\right)^2 \right\} \:+\:\phi_{0}\,. \label{PowerK0NLSSol} \nonumber \\
\eea
When $q = 2/n$ or $n =0$, the field has infinite value.  $q$ and
$\epsilon$ must have the same sign for the solution to be real. The
last logarithmic term does not restrict sign of $q$. This is unlike
the solution obtained from Friedmann formulation which requires
$q<0$ which violates power-law expansion condition ($q>1$). Working
in neither of them can obtain exact solution with $k\neq 0$. In NLS
formulation, we can not set $D$ to zero while $n$ is multiplied to
the other terms then it can not be reduced to the scalar dominant
case. This is a weak aspect. Obviously, the most difficult case is
when $k\neq 0$ with $D\neq 0$. This case can not be integrated out
in both frameworks unless assuming $n=2$ (equivalent to $w_{\gamma}
= -1/3$) which is not physical.

There are other good aspects of the NLS formulation. Since
transforming standard Friedmann formulation ($t$ as independent
variable) to NLS formulation ($x$ as independent variable) makes
 $n$ appear in all terms of the integrand and also changes fluid
 density
 term $D$ from time-dependent term to a constant $E$,
therefore the number of $x$ (or equivalently $t$)-dependent terms is
reduced by one and hence simplifying the integral (\ref{phitoPx}).
In the case of exponential (de Sitter) expansion using NLS
formulation, the solution when $k\neq 0$ and $D\neq 0$ can be
obtained without assuming $n$ value but $n=0, 2, 3, 4, 6$ must be
given if working within Friedmann formulation. The phantom expansion
case is very similar to the power-law case but only with different
sign (see Ref. \cite{Phetnora:2008}).

\section{Phantom cosmology and Big Rip singularity} \label{sec:BigRip}
\begin{figure}[t]
\begin{center}
\includegraphics[width=5.4cm,height=5.4cm,angle=0]{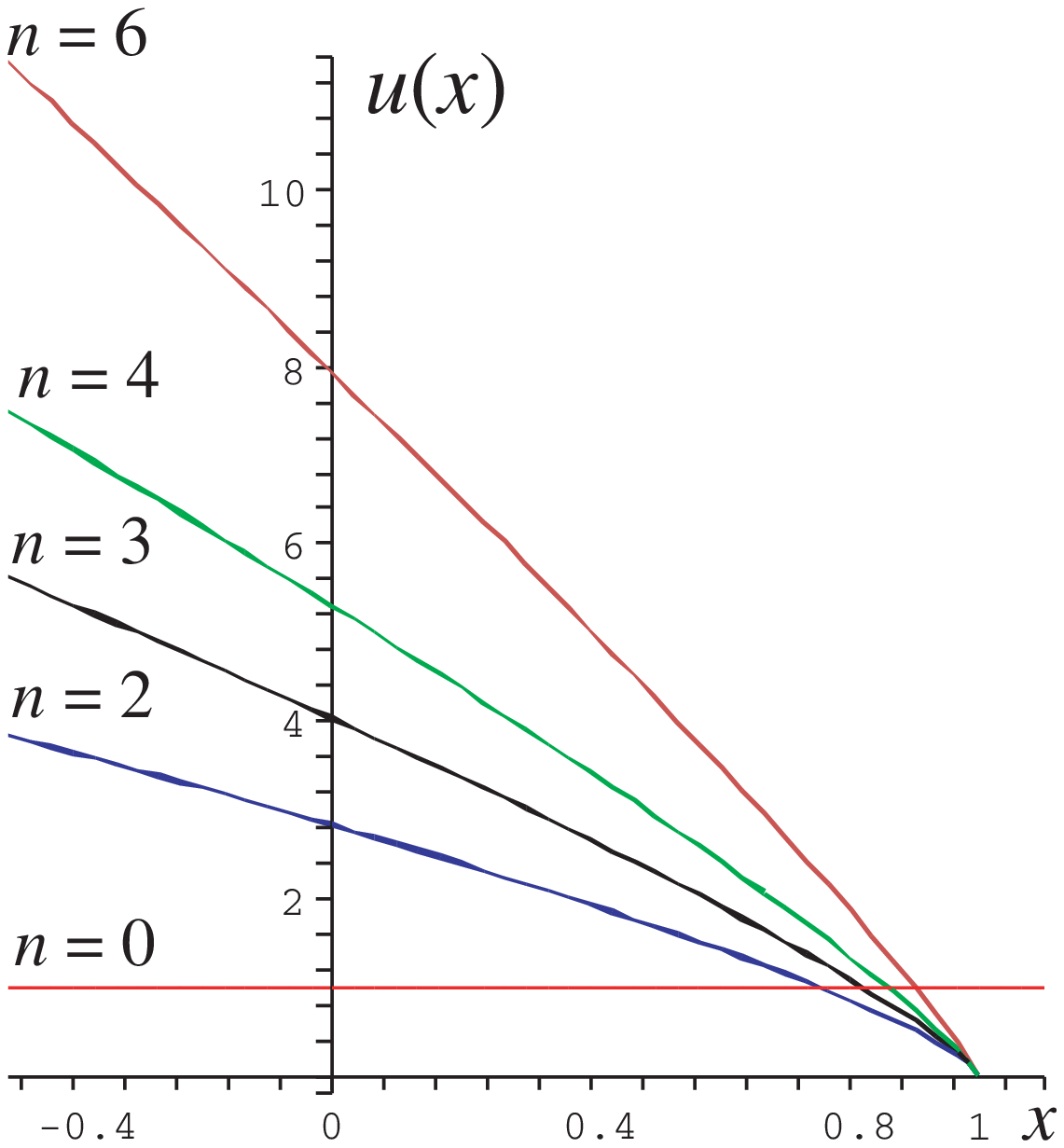}
\end{center}
\caption{Schr\"{o}dinger wave function, $u(x)$ when assuming phantom
expansion. $u(x)$ depends on only $q$, $n$  and $t_{\rm a}$. Here we
set $t_{\rm a} =1.0$ and $q = -6.666$. If $k=0$, $q=-6.666$
corresponds to $w_{\rm eff} = -1.1$. \label{figux}}
\begin{center}
\includegraphics[width=7.5cm,height=9.6cm,angle=0]{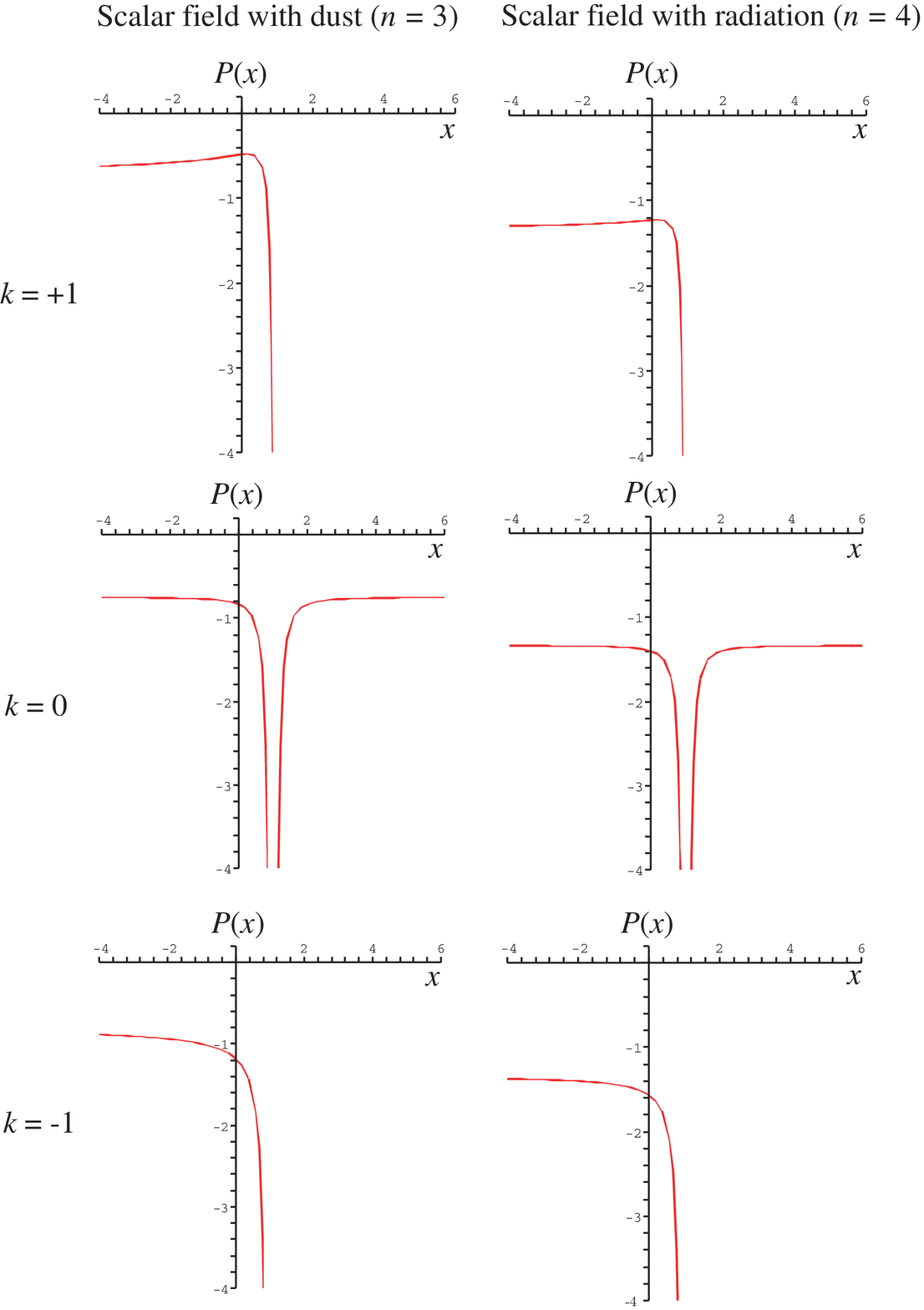}
\end{center}
\caption{Schr\"{o}dinger potential in phantom expansion case for
dust and radiation fluids with $k= 0, \pm1$. Numerical parameters
are as in the $u(x)$ plots (Fig. \ref{figux}). $x_0$ is set to $1$.
For non-zero $k$, there is only one real branch of $P(x)$.
\label{figpx}}
\end{figure}

If we assume the expansion to a form, $a(t) \sim (t_{\rm a} - t)^q$
with a finite time $t_{\rm a}$, one can see that $q = 2/3(1+w_{\rm
eff}) < 0$ (for a flat universe). This corresponds to $w_{\rm eff} <
-1$. Such equation of state is called phantom. The Schr\"{o}dinger
scale, $x$ is related to cosmic time scale, $t$ as \bea x(t)  =
\frac{1}{\beta}\left[ (t_{\rm a} - t)^{-\beta}\right] + x_0\,,
\label{xast} \eea and the wave function is
\bea u(x)  =  \left[ \beta (x-x_0) \right]^{qn/(qn-2)}\,%
\eea
which is plotted in Fig. \ref{figux} with various types of
barotropic fluid \cite{Phetnora:2008}. Therefore \bea
 P(x) &=& \frac{2qn}{(qn-2)^2}\frac{1}{(x-x_0)^2}
 + \frac{kn}{2}\left[ \frac{2}{(qn-2)(x-x_0)}
\right]^{2q(n-2)/(qn-2)}  -  \frac{\kappa^2 n^2 D }{12}.
\label{Pxphantom} \eea
 Fig. \ref{figpx} shows $P(x)$ plots for three
cases of $k$ with dust and radiation. $P(x)$ goes to negative
infinity at $x = x_0 = 1$.
Expansion of the form, $a(t) \sim (t_{\rm a} - t)^q$ leads to unwanted future Big Rip singularity \cite{Caldwell:1999ew}. The Big Rip conditions
are that $(a, \rho_{\rm tot}, |p_{\rm tot}|) \rightarrow \infty$ which happen when $t\rightarrow t_{\rm a}^{-}$ in finite future time. Written
in NLS language, if $a \rightarrow \infty$, $x\rightarrow x_0^{-}$ and then $u\rightarrow 0^{+}$ (see Fig. \ref{figux}). Considering also Eqs.
(\ref{rhoNLS}) and (\ref{pNLS}), hence conditions of the Big Rip singularity are \cite{Gumjudpai:2008}
\bea
t\rightarrow t_{\rm a}^{-}\:\;\;\; &\Leftrightarrow&\;\;\;\;\;\;\;\;\;  x \rightarrow x_0^{-} \no \\
a \rightarrow \infty \;\;\; &\Leftrightarrow&\;\;\;\; u(x) \rightarrow 0^{+}  \no \\
\rho_{\rm tot} \rightarrow \infty\;\;\; &\Leftrightarrow&\;\;\;  u'(x) \rightarrow \infty  \no \\
|p_{\rm tot}| \rightarrow \infty\;\;\; &\Leftrightarrow&\;\;\;
u'(x) \rightarrow \infty\,. \label{NLSBigRip} \eea
We have one less infinite value in NLS Big Rip condition, i.e. $u(x)$ goes to zero. The NLS effective equation of state $w_{\rm eff} = p_{\rm
tot}/\rho_{\rm tot}$ can be expressed using Eqs. (\ref{rhoNLS}) and (\ref{pNLS}). Approaching the Big Rip, $x \rightarrow x_0^{-}, u \rightarrow
0^{+}$, then $w_{\rm eff} \rightarrow -1 + {2}/{3q}$ where $q<0$ is a constant. This limit is the same as the effective phantom equation of
state in the case $k=0$. It is important to note that scalar field potential here is built phenomenologically based on expansion function, not
on fundamental physics.

\section{WKB Approximation} \label{sec:WKB}
WKB approximation in quantum mechanics is a tool to obtain wave
function. However, in NLS formulation of scalar field cosmology, the
wave function is first presumed before working out the shape of
$P(x)$. Procedure is opposite to that of quantum mechanics. Hence
the WKB approximation might not be needed at all for the NLS.
Anyway, if one wants to test the WKB approximation in the NLS
formulation, these below are some results. The WKB are valid when
the coefficient of highest-order derivative term in the
Schr\"{o}dinger equation is small or when the potential is very
slowly-varying. Consider linear case of Eq. (\ref{schroeq}),
($k=0$),
 \bea - u'' + \l[{P}(x) -{E}\r] u = 0 \,.
\label{WKB1} \eea In Figs \ref{pic4} and \ref{figpx}, the left-hand
side of $P(x)$ is physical since it corresponds to positive time. In
most regions, there are ranges of slowly varying $P(x)$ at large
value of $|x|$, in which the WKB is valid. The approximation gives
\bea a \sim A \exp\l[{\pm (2/n) i \int_{x_1}^{x_2} \sqrt{E-P(x)} }
\,{\rm d} x \r]\,, \label{NLSWKBa} \eea where $A$ is a constant.

\section{Conclusions} \label{sec:con}
Here we conclude aspects of NLS-type formulation of scalar field
cosmology. The NLS-type formulation is well-applicable in presence
of barotropic fluid and a canonical scalar field. There are few
advantages of the NLS formulation as well as disadvantages to the
conventional Friedmann formulation. With hope that some more
interesting and useful features could be revealed in future.

\section*{Acknowledgments} The author thanks Jeong Ryeol Choi for invitation to write this review.
The author is a TRF Research Scholar under the Thailand Research Fund. In Cambridge, the author is supported by Naresuan University's Overseas
Visiting Postdoctoral Research Fellowship and the Centre for Theoretical Cosmology, D.A.M.T.P., University of Cambridge at which very much
gratitude is expressed to Anne-Christine Davis, Stephen Hawking and Neil Turok for providing the centre's subsistence.


\begin{thebibliography}{99}


\bibitem{Masi:2002hp}
  S.~Masi {\it et al.},
  Prog.\ Part.\ Nucl.\ Phys.\  {\bf 48}, 243 (2002).

\bibitem{Scranton:2003in}
 R.~Scranton {\it et al.} (SDSS Collaboration),
arXiv: astro-ph/0307335.


\bibitem{Riess:1998cb}
 A.~G.~Riess {\it et al.} (Supernova Search Team Collaboration),
   Astron.\ J.\  {\bf 116}, 1009 (1998);
S.~Perlmutter {\it et al.} (Supernova Cosmology Project
Collaboration),
   Astrophys.\ J.\  {\bf 517}, 565 (1999);
A.~G.~Riess,
arXiv: astro-ph/9908237;
G.~Goldhaber {\it et al.}  (The Supernova Cosmology Project
Collaboration),
Astrophys.\ J., {\bf 558}, 359 (2001); 
J.~L.~Tonry {\it et al.} (Supernova Search Team Collaboration),
   Astrophys.\ J.\ {\bf 594}, 1 (2003).
\bibitem{Riess:2004nr}
  A.~G.~Riess {\it et al.} (Supernova Search Team Collaboration),
   Astrophys.\ J.   {\bf 607}, 665 (2004);
   A.~G.~Riess {\it et al.},
     Astrophys.\ J. {\bf 659}, 98 (2007);
\bibitem{Astier:2005qq}
   P.~Astier {\it et al.} (SNLS Collaboration),
   Astron.\ Astrophys.   {\bf 447}, 31 (2006).

\bibitem{Dvali}
  G. R. Dvali, G. Gabadadze and M. Porrati, Phys. Lett. B {\bf 485}, 208 (2000); B.~Gumjudpai,
  Gen.\ Rel.\ Grav.\  {\bf 36}, 747 (2004); R. Maartens, Living Rev. Rel. {\bf 7}, 7 (2004);
  S.~Nojiri and S.~D.~Odintsov,
  Int.\ J.\ Geom.\ Meth.\ Mod.\ Phys.\  {\bf 4}, 115 (2007).


\bibitem{Kolb:2005me}
  E.~W.~Kolb, S.~Matarrese, A.~Notari and A.~Riotto,
  arXiv:hep-th/0503117;
  E.~W.~Kolb, S.~Matarrese and A.~Riotto,
  New J.\ Phys.\  {\bf 8}, 322 (2006).

\bibitem{Padmanabhan:2004av}
  T.~Padmanabhan,
  Curr.\ Sci.\  {\bf 88}, 1057 (2005);
  E.~J.~Copeland, M.~Sami and S.~Tsujikawa,
  Int.\ J.\ Mod.\ Phys.\  D {\bf 15}, 1753 (2006);
  T.~Padmanabhan,
  AIP Conf.\ Proc.\  {\bf 861}, 179 (2006).

\bibitem{inflation}
D.~Kazanas,
  Astrophys.\ J.\  {\bf 241}, L59 (1980);
A. A. Starobinsky,  Phys. Lett. B {\bf 91}, 99 (1980); A. H. Guth, Phys. Rev. D {\bf 23}, 347 (1981);
               K. Sato,  Mon. Not. Roy. Astro. Soc.  {\bf 195}, 467
               (1981);
                 A. Albrecht and P. J. Steinhardt,  Phys. Rev. Lett. {\bf 48}, 1220 (1982);
                 A. D. Linde,  Phys. Lett. B {\bf 108}, 389
                 (1982).


\bibitem{Hinshaw:2008kr}
 D.~N.~Spergel {\it et al.}  (WMAP Collaboration),
  Astrophys.\ J.\ Suppl.\  {\bf 170}, 377 (2007).

\bibitem{Caldwell:1999ew}
  R.~R.~Caldwell,
  Phys.\ Lett.\ B {\bf 545}, 23 (2002)
  G.~W.~Gibbons,
  arXiv:hep-th/0302199;
  S.~Nojiri and S.~D.~Odintsov,
  Phys.\ Lett.\  B {\bf 562}, 147 (2003).




\bibitem{Melchiorri:2002ux}
A.~Melchiorri, L.~Mersini-Houghton, C.~J.~Odman and M.~Trodden,
  { Phys.\ Rev.\ D} {\bf 68}, 043509 (2003);
  P.~S.~Corasaniti, M.~Kunz, D.~Parkinson, E.~J.~Copeland and B.~A.~Bassett,
  { Phys.\ Rev.\  D} {\bf 70}, 083006 (2004);
  U.~Alam, V.~Sahni, T.~D.~Saini and A.~A.~Starobinsky,
  { Mon.\ Not.\ Roy.\ Astron.\ Soc.}  {\bf 354}, 275 (2004).

\bibitem{Dunkley:2008ie}
 G.~Hinshaw {\it et al.} (WMAP Collaboration),
  arXiv:0803.0732 [astro-ph];
  J.~Dunkley {\it et al.} (WMAP Collaboration),
  arXiv:0803.0586 [astro-ph];
 E.~Komatsu {\it et al.} (WMAP Collaboration),
  arXiv:0803.0547 [astro-ph].








\bibitem{Percival:2007yw}
  W.~J.~Percival, S.~Cole, D.~J.~Eisenstein, R.~C.~Nichol, J.~A.~Peacock, A.~C.~Pope and A.~S.~Szalay,
  { Mon.\ Not.\ Roy.\ Astron.\ Soc.}  {\bf 381}, 1053 (2007).




\bibitem{WoodVasey:2007jb}
   W.~M.~Wood-Vasey {\it et al.} (ESSENCE Collaboration),
  { Astrophys.\ J.}  {\bf 666}, 694 (2007).




\bibitem{Steigman:2007xt}
  G.~Steigman,
  { Ann.\ Rev.\ Nucl.\ Part.\ Sci.}  {\bf 57}, 463 (2007).
\bibitem{Wright:2007vr}
  E.~L.~Wright,
  { Astrophys.\ J. }  {\bf 664}, 633 (2007).
\bibitem{Caldwell:2003vq}
A. A. Starobinsky, Grav. Cosmol. {\bf 6}, 157  (2000);
  R.~R.~Caldwell, M.~Kamionkowski and N.~N.~Weinberg,
  { Phys.\ Rev.\ Lett.}  {\bf 91}, 071301  (2003);
   J.~g.~Hao and X.~z.~Li,
  { Phys.\ Rev.\ D} {\bf 67}, 107303  (2003);
  S.~Nesseris and L.~Perivolaropoulos,
  { Phys.\ Rev.\ D } {\bf 70}, 123529  (2004);
   X.~z.~Li and J.~g.~Hao,
  { Phys.\ Rev.\ D} {\bf 69}, 107303 (2004);
  J.~G.~Hao and X.~z.~Li,
  { Phys.\ Rev.\ D } {\bf 70}, 043529 (2004);
 M.~Sami and A.~Toporensky,
  { Mod.\ Phys.\ Lett.\ A} {\bf 19}, 1509 (2004);
S.~Nojiri, S.~D.~Odintsov and S.~Tsujikawa,
  Phys.\ Rev.\  D {\bf 71}, 063004 (2005);
  B.~Gumjudpai, T.~Naskar, M.~Sami and S.~Tsujikawa,
  { J. Cosmol. Astropart. Phys.} {\bf 0506}, 007 (2005);
  L.~A.~Urena-Lopez,
   { J. Cosmol. Astropart. Phys.}  {\bf 0509}, 013 (2005).
\bibitem{Sami:2005zc}
P.~Singh, M.~Sami and N.~Dadhich,
  { Phys.\ Rev.\ D} {\bf 68}, 023522 (2003);
  S.~Nojiri, S.~D.~Odintsov and M.~Sasaki,
  { Phys.\ Rev.\  D} {\bf 71}, 123509 (2005);
  M.~Sami, A.~Toporensky, P.~V.~Tretjakov and S.~Tsujikawa,
  {  Phys.\ Lett.\  B} {\bf 619}, 193 (2005);
  G.~Calcagni, S.~Tsujikawa and M.~Sami,
  { Class.\ Quant.\ Grav. }  {\bf 22}, 3977 (2005);
  H.~Wei and R.~G.~Cai,
  { Phys.\ Rev.\ D} {\bf 72}, 123507 (2005);
   P.~X.~Wu and H.~W.~Yu,
  Nucl.\ Phys.\  B {\bf 727}, 355 (2005);
  T. Koivisto and D. F. Mota,
 { Phys. Lett. B} {\bf 644}, 104 (2007);
T. Koivisto and D. F. Mota,
 { Phys. Rev. D} {\bf 75}, 023518 (2007);
  B.~M.~Leith and I.~P.~Neupane,
  J. Cosmol. Astropart. Phys. {\bf 0705}, 019 (2007);
  D.~Samart and B.~Gumjudpai,
  { Phys. Rev. D} {\bf 76}, 043514 (2007); T. Naskar and J. Ward, { Phys. Rev. D} {\bf 76}, 063514 (2007); B.~Gumjudpai,
  { Thai J. Phys. Series 3: Proc. of the SIAM Phys. Cong. 2007}, [arXiv:0706.3467 [gr-qc]].


\bibitem{carroll} S. M. Carroll, M. Hoffman and M. Trodden, Phys. Rev. D {\bf 68}, 023509 (2003).







\bibitem{Hawkins:2001zx}
  R.~M.~Hawkins and J.~E.~Lidsey,
  Phys.\ Rev.\ D {\bf 66}, 023523 (2002);
  F.~L.~Williams and P.~G.~Kevrekidis,
  Class.\ Quant.\ Grav.\  {\bf 20}, L177 (2003);
  F.~L.~Williams, P.~G.~Kevrekidis, T.~Christodoulakis, C.~Helias, G.~O.~Papadopoulos and T.~Grammenos,
 Trends in Gen. Rel. and Quant. Cosmol., Nova
Science Pub. 37-48 (2006). 
\bibitem{Williams:2005bp}
  F.~L.~Williams,
  Int.\ J.\ Mod.\ Phys.\  A {\bf 20}, 2481 (2005).




\bibitem{Lidsey:2003ze}
  J.~E.~Lidsey,
  Class.\ Quant.\ Grav.\  {\bf 21}, 777 (2004).

\bibitem{Kamenshchik2006}
 A.~Kamenshchik, M.~Luzzi and G.~Venturi, arXiv: math-ph/0506017.


\bibitem{D'Ambroise:2006kg}
  J.~D'Ambroise and F.~L.~Williams,
  Int. J. Pure Appl. Maths. {\bf 34}, 117 (2007).



\bibitem{Gumjudpai:2007qq}
  B.~Gumjudpai,
  Astropart. Phys. {\bf 30}, 186 (2008).


\bibitem{Gumjudpai:2007bx} B.~Gumjudpai,
 Gen. Rel. Grav. {\bf 41}, 249 (2009).  

\bibitem{Phetnora:2008} T. Phetnora, R. Sooksan and B.~Gumjudpai,
arXiv: 0805.3794 [gr-qc].
\bibitem{Gumjudpai:2008} B.~Gumjudpai, J. Cosmol. Astropart. Phys. {\bf 0809}, 028 (2008).

\bibitem{Chervon:1999}
  S. V. Chervon and V. M. Zhuravlev,
  arXiv: gr-qc/9907051; A. V. Yurov, arXiv: astro-ph/0305019; A. A.
  Andrianov, F. Cannata and A.  Kamenshchik, Phys. Rev. D {\bf
  72}, 043531  (2005); A. V. Yurov, A. V. Astashenok and V. A. Yurov, arXiv: astro-ph/0701597.



\bibitem{liddlebook} A.~R.~Liddle and D.~H.~Lyth, Cosmological
Inflation and Large-Scale Structure, Cambridge University Press
(2000).


\bibitem{Liddle:1992wi} A.~R.~Liddle and D.~H.~Lyth,
  Phys.\ Lett.\  B {\bf 291}, 391 (1992).  


\bibitem{Chaicherdsakul:2006ui}
  K.~Chaicherdsakul,
  Phys.\ Rev.\ D\  {\bf 75}, 063522 (2007).
\bibitem{Lucchin} F. Lucchin and S. Matarrese, Phys. Rev. D {\bf 32}, 1316 (1985).
\bibitem{Zhu:2007tm}
  Z.~H.~Zhu, M.~Hu, J.~S.~Alcaniz and Y.~X.~Liu,
  Astron. and Astrophys. {\bf 483}, 15 (2008)
\bibitem{tepsuriya:2008}
K. Tepsuriya and B. Gumjudpai, in preparation. 

\end{thebibliography}
\end{document}